%% file: article.tex
\documentclass[pdflatex,sn-mathphys]{config/sn-jnl}

\usepackage{academicons}

\usepackage{orcidlink}
\definecolor{orcidlogocol}{HTML}{A6CE39}
\usepackage{url}

\usepackage[acronym,toc]{glossaries}
\usepackage{colortbl,hhline}
\usepackage{float}
\usepackage{comment}
\loadglsentries{glossary}
\makeglossaries
\usepackage{soul}

\jyear{2023}%

\theoremstyle{thmstyleone}%
\theoremstyle{thmstyletwo}%

\theoremstyle{thmstylethree}%

\begin{document}

\title[Secure Video Streaming Using Dedicated Hardware]{Secure Video Streaming Using Dedicated Hardware}


\author[1]{\fnm{Nicholas} \sur{Murray-Hill}}\email{n0821510@my.ntu.ac.uk}

\author[1]{\fnm{Laura} \sur{Fontes}\orcidlink{0000-0003-0171-7436}}\email{n1119003@my.ntu.ac.uk}

\author*[1]{\fnm{Pedro} \sur{Machado}\orcidlink{0000-0003-1760-3871}}\email{pedro.machado@ntu.ac.uk}

\author[1]{\fnm{Isibor Kennedy} \sur{Ihianle}\orcidlink{0000-0001-7445-8573}}\email{isibor.ihianle@ntu.ac.uk}

\affil*[1]{\orgdiv{Department of Computer Science, School of Science and Technology}, \orgname{Nottingham Trent University}, \orgaddress{\street{ERD220, Clifton Campus}, \city{Nottingham}, \postcode{NG11 8NS}, \state{Nottinghamshire}, \country{UK}}}




\abstract{The purpose of this article is to present a system that enhances the security, efficiency, and reconfigurability of an \gls*{iot} system used for surveillance and monitoring. A \gls*{mpsoc} composed of \gls*{cpu} and \gls*{fpga} is proposed for increasing the security and the frame rate of a smart \gls*{iot} edge device. The private encryption key is safely embedded in the \gls*{fpga} unit to avoid being exposed in the \gls*{ram}. This allows the edge device to securely store and authenticate the key, protecting the data transmitted from the same \gls*{ic}. Additionally, the edge device can simultaneously publish and route a camera stream using a lightweight communication protocol, achieving a frame rate of 14 \gls*{fps}. The performance of the \gls*{mpsoc} is compared to a \gls*{njn} and a \gls*{rpi4} and it is found that the \gls*{rpi4} is the most cost-effective solution but with lower frame rate, the \gls*{njn} is the fastest because it can achieve higher frame-rate but it is not secure, and the  \gls*{mpsoc} is the optimal solution because it offers a balanced frame rate and it is secure because it never exposes the secure key into the memory. The proposed system successfully addresses the challenges of security, scalability, and efficiency in an \gls*{iot} system used for surveillance and monitoring. The \gls*{rsa} encryption key is securely stored and authenticated, and the edge device is able to simultaneously publish and route a camera stream feed high-definition images at 14 \gls*{fps}. The proposed system enhances the security, efficiency, and reconfigurability of an \gls*{iot} system used for surveillance and monitoring. The \gls*{rsa} encryption key is effectively protected by the implementation of an \gls*{mpsoc} with \gls*{pl}, which also allows for faster processing and data transmission.}

\keywords{\gls*{iot}, \gls*{fpga}, \gls*{gpu}, PYNQ, \gls*{mpsoc}, security streaming, edge computing}



\maketitle

\section{Introduction} \label{Ch:introduction}
\acrfull*{iot} has become a central focus in the technology industry, with the development of smart devices that collect and exchange data over the internet. These devices, including sensors, actuators, and other \gls*{iot}-enabled technologies, are used for a variety of purposes such as analysis, processing, and automation. It is estimated that there will be approximately 75 billion \gls*{iot} devices in use by 2025 \cite{nord2019}. The use of \gls*{iot} technology has expanded into various fields including smart energy, industrial factories, transportation, and home automation.

However, the widespread integration of \gls*{iot} systems in our daily lives also leads to an increase in the amount of data being collected and exchanged over the internet, raising concerns about scalability and the protection of sensitive and private data. Many \gls*{iot} systems rely on centralised architectures, which process and perform security operations on the cloud \cite{kouicem2018}. These architectures have limitations such as scalability, transaction speeds, interoperability, and privacy/security \cite{farahani2021}, and may also be a single point of failure in the event of a breach. Centralised servers that manage keys and act as a single trust authority can pose a significant security risk to other devices on the network, as these keys often form the foundation of security systems, cryptography algorithms, and device authentication/verification \cite{kouicem2018}.

To address these issues, distributed infrastructures have emerged, allowing modern \gls*{iot} edge devices to perform their own processing and transmit data at the edge. These devices feature robust security mechanisms for secure authentication and secret key storage, which can be used for encryption in a secure, hardware-enforced manner to improve end-device security for \gls*{iot} systems and data integrity \cite{johnson2015}. \acrfull*{fpga} offer such security features through the use of programmable logic and reconfigurability after manufacturing. This reconfigurability allows devices to be updated in order to keep up with the constantly evolving technology landscape and emerging security threats.

The main contribution of this article is the development and implementation of a secure, efficient, and reconfigurable surveillance and monitoring \gls*{iot} system using dedicated hardware. By utilising the \acrfull*{mpsoc}'s \acrfull*{pl} to securely store and authenticate a symmetric key, the proposed system improves the security and integrity of data transmitted from an edge device. Additionally, the use of \glspl*{mpsoc} allow the edge device to simultaneously publish and route a camera stream using a lightweight communication protocol, achieving a high capture rate. The proposed system addresses many of the challenges faced by current \gls*{iot} systems, including scalability, security, and efficiency.

The remainder of this article is organised as follows: relevant literature and related works to the proposed \gls*{iot} system proposed in this article is reviewed in Section \ref{Ch:lit_rev}, the methodology used to develop and implement the proposed system is described in Section \ref{Ch:methods}, the results and analysis of the proposed system are presented in Section \ref{Ch:results}, and finally, the conclusion and future work are discussed in Section \ref{Ch:concl_future}.

\section{Related Work}\label{Ch:lit_rev}
The \gls*{iot} is a technology that aims to provide an infrastructure for applications that can coordinate the interaction of people, things, and systems for a specific purpose \cite{wortmann2015}. These applications do not necessarily have a universally adopted standard, but the architectural model of an \gls*{iot} system typically consists of three main layers: the perception layer, which uses sensors and microcontrollers to perceive the physical environment; the communication or network layer, which processes and transports data; and the application layer, which uses the data to deliver application-specific services to the user \cite{nord2019}. The communication/network layer uses various technologies to package and transmit data due to the processing and bandwidth restrictions of many \gls*{iot} devices. \gls*{http}, which is commonly used for communication between devices on the internet, is not suitable for low-powered \gls*{iot} devices due to its fully connection-oriented architecture, large header size, and latency \cite{patel2020, ref1}. Established communication protocols that are more suitable for these power and bandwidth-restricted \gls*{iot} requirements include \gls*{coap}, \gls*{mqtt}, and \gls*{xmpp}. Corak et al. \cite{corak2018} evaluated and compared the performance of these protocols in a real-world \gls*{iot} testbed. The metrics considered were packet creation time and packet delivery speed to determine the delay differences. The study found that \gls*{xmpp} had the worst performance due to its use of \gls*{xml} format, which increased latency. \gls*{mqtt} and \gls*{coap} had similar overall performance in terms of packet creation and transmission time, but \gls*{mqtt} was found to be more optimised and standardised. In addition to these protocols, wireless technologies such as \gls*{lora}, \gls*{lorawan}, and \gls*{lpwan} can be used to enable long-range and low-power communications for \gls{iot} devices \cite{queralta2019}. These wireless protocols are designed to provide low-power, wide-area networks, making them ideal for use cases where devices need to transmit small amounts of data over long distances, such as in agriculture, smart cities, and industrial applications \cite{queralta2019}. Van der Westhuizen and Hancke  \cite{van2018} conducted a more in-depth comparison between \gls*{coap} and \gls*{mqtt} to determine which was the most suitable for use with constrained devices, specifically sensors. The comparison considered communication delay and network traffic. Both protocols were found to be good choices for resource-constrained devices, with similar performance and response times. However, the most suitable protocol depended on the overall requirements of the system. \gls*{coap} was found to be the optimal choice for interfacing with business systems, due to its small average packet sizes and minimal battery/data usage. \gls*{mqtt}, on the other hand, was found to be the preferred solution for systems such as home automation and sensor networks, where device heterogeneity is more pronounced. \gls*{mqtt} was easier to configure for new devices and had the most effective data flow thanks to its publish/subscribe model and use of \gls*{qos}.

\subsection{Edge computing}
Edge computing is a network architecture that involves processing sensory (e.g. visual data) data closer to the source, rather than on the cloud \cite{khan2019}. This allows for fast processing and efficient handling of data intensive operations in real-world scenarios such as the \gls*{iot}. While edge computing can offer benefits for \gls*{iot} systems, there are also limitations in terms of security. Khan et al. \cite{khan2019, ref2} found that further development is needed in areas such as authentication and access control, and that tamper-proof architectures may be one solution to addressing these security issues. However, securing large scale and time-critical \gls*{iot} systems can also be challenging due to the cost of methods such as encryption in terms of latency, energy consumption, and network bandwidth \cite{mohanty2020}. Additionally, the heterogeneity of devices that communicate across these networks without a well-established protocol can also pose challenges \cite{nord2019}. Fortunately, professionals in the field are working to overcome these limitations and improve the safety and efficiency of communication between \gls*{iot} devices.

\subsection{FPGA technology}
\glspl*{fpga} are specialised hardware devices that have gained popularity in the edge computing space due to their ability to solve problems through reconfigurable hardware circuits. These circuits can be described using \gls*{hdl} such as Verilog and \gls*{vhdl}, and are made up of various logic units such as look-up tables, flip-flops, and multiplexers. \glspl*{fpga} offer several benefits for security, parallel computing, and flexibility to update hardware designs after deployment \cite{elnawawy2019, ref3, ref4}. They have also been advanced through the use of \gls*{soc}, which integrate programmable logic with real-time processors. An example of this is the AMD-Xilinx Zynq Ultrascale+ \gls*{mpsoc}\footnote{Available online, \protect\url{https://www.xilinx.com/products/silicon-devices/soc/zynq-ultrascale-mpsoc.html}, last accessed 07/01/2023}, which includes an \gls*{arm} \gls*{cpu}, programmable logic, and units for graphics and video processing. While \glspl*{fpga} and \glspl*{soc} have similarities with microcontrollers, \glspl*{fpga} offer advantages in physical and cybersecurity through encrypted bitstreams and key loading mechanisms, and can act as a \gls*{rot} by holding security private keys and critical algorithms. \glspl*{fpga} also show greater efficiency in processing algorithms for image processing and video transcoding due to their parallel computing capabilities.

While \glspl*{fpga} offer significant advantages for \gls*{iot}, they are considered complex due to the low-level hardware knowledge required, such as \gls*{vhdl} and Verilog. To address this, \gls*{fpga} vendors have been promoting the use of high-level design flows and tools that allow for the creation of \gls*{rtl} designs using high-level languages like C, C++, System C and \gls*{opencl}. However, the question remains as to how well these high-level designs compare to manually written \gls*{rtl} designs in terms of optimisation. Guo et al. \cite{guo2017} discussed that while \gls*{hls} may not be as optimised as manually written \gls*{rtl} designs for complex designs, the use of directives like loop unrolling and loop merging and pipelining can significantly improve resource utilisation, reduce latency, increase resource sharing, and optimize logic for video processing algorithms. These findings suggest that \gls*{fpga} technology can be more accessible to designers without strong low-level hardware knowledge, while still maintaining good performance.

In summary, the reviewed articles have demonstrated the various considerations and challenges faced in the design and implementation of an \gls*{iot} system. Communication protocols such as \gls*{mqtt} and \gls*{coap} have been shown to be effective in resource constrained environments, but the choice between them ultimately depends on the specific requirements of the system. Edge computing has the potential to improve the efficiency and security of \gls*{iot} networks, but also comes with its own limitations that require further development. \gls*{fpga} technology offers advanced security and parallel processing capabilities for \gls*{iot}, but can be complex to implement. High level synthesis tools, such as the AMD-Xilinx Vivado \gls*{hls}\footnote{Available online, \protect\url{https://www.xilinx.com/support/documentation-navigation/design-hubs/dh0090-vitis-hls-hub.html}, last accessed 07/01/2023}, have been shown to improve the productivity and performance of \gls*{fpga} designs for real time image processing applications, but may not always be as optimised as manually written designs. These findings highlight the importance of carefully evaluating the various technologies and approaches available for a particular \gls*{iot} system in order to ensure optimal performance and security.

\section{Methods}\label{Ch:methods}
The proposed \gls*{iot} system utilises the \acrfull*{ultra96} equipped with a powerful AMD-Xilinx Zynq UltraScale+ \gls*{mpsoc} ZU3EG\footnote{Available online, \protect\url{https://www.xilinx.com/content/dam/xilinx/imgs/products/zynq/zynq-eg-block.PNG}, last accessed 07/01/2023} device as the main processing system at the perception layer. The performance of the \gls*{ultra96} was compared to a \acrfull*{njn} and a \acrfull*{rpi4} under the same testing conditions. 

\begin{figure}[] 
     \centering
     \includegraphics[scale=0.5]{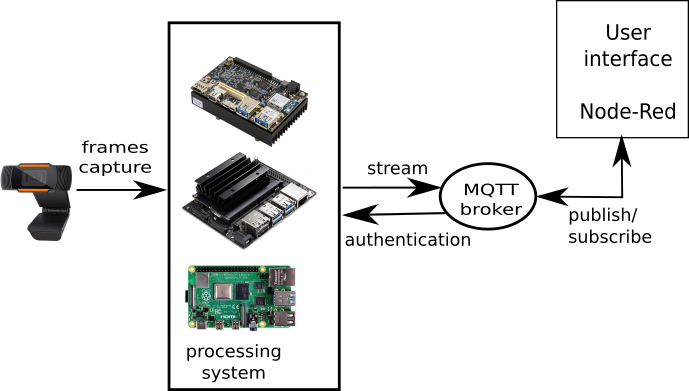}
     \caption{Architecture Diagram}
     \label{fig:arch}
\end{figure}

To establish a fair comparison, each processing device (i.e. \gls*{ultra96}, \gls*{rpi4} and \gls*{njn}) runs an \gls*{mqtt} client to publish data from its connected \gls*{usb} webcam to an \gls*{mqtt} broker, which acts as an intermediary to route the data to interested parties. The camera feed is then displayed on a Node-RED\footnote{Available online, \protect\url{https://nodered.org/}, last accessed 07/01/2023} dashboard at the application layer for subscribers to view. The use of \gls*{mqtt} and the Node-RED dashboard allows for efficient and flexible communication and data management within the system. The system also implements security measures, such as bitstream authentication, to protect against potential attacks. Overall, the proposed \gls*{iot} system utilises a variety of technologies to coordinate the interaction of people, things, and systems for a specific purpose (see Figure~\ref{fig:arch}).

The Avnet \gls*{ultra96}, which is powered by an AMD-Xilinx Zynq UltraScale+ \gls*{mpsoc} device that combines an \gls*{arm} processor and \gls*{fpga}. The \gls*{ultra96} is energy efficient and performs well due to designated processors being responsible for specific tasks. The \gls*{ps} in the AMD-Xilinx Zynq UltraScale+ \gls*{mpsoc} runs an \gls*{arm}64v8 Linux environment for running a web server while also interfacing with the programmable logic via the \gls*{axi4} for user authentication and on-field reconfigurability. \gls*{arm}64v8 is a version of the \gls*{arm} architecture that supports 64-bit instructions. It is used in some 64-bit \gls*{arm} processors, such as those used in the Avnet \gls*{ultra96}.

The \gls*{njn}\footnote{Available online, \protect\url{https://developer.nvidia.com/embedded/jetson-nano-developer-kit}, last accessed 15/01/2023} is a small, powerful computer designed for use in image and video processing applications. It is powered by a quad-core \gls*{arm} Cortex-A57 \gls*{cpu} and a 128-core NVIDIA Maxwell \gls*{gpu}, running on an \gls*{arm}64v8 Linux environment. Programming the \gls*{njn} is typically done using the \gls*{njpsdk}, which includes a Linux-based development environment and a variety of software libraries for working with the \gls*{cpu} and \gls*{gpu}. \gls*{opencv} compiled with the CUDA library was used to achieve maximum performance when processing visual data.

The \gls*{rpi4}\footnote{Available online, \protect\url{https://www.raspberrypi.com/products/raspberry-pi-4-model-b/}, last accessed {15/01/2023}} is a low-cost, single-board computer designed for educational and hobbyist use. It is powered by a Broadcom BCM2711, quad-core Cortex-A72 (\gls*{arm}v8) 64-bit SoC @ 1.5GHz and runs on a Linux-based operating system, typically Raspbian. \gls*{opencv} was used to achieve maximum performance when processing visual data. The \gls*{rpi4} is known to be cost-effective solution but with lower frame rate compared to other devices.

The \gls*{iot} system was not initially configured with any software or input sensors, so a USB webcam was connected to the \gls*{usb} 3.0 Type A port to capture the camera feed. The Micro-B upstream port was used to connect to a host workstation on the same local network, although the board also supports WiFi connectivity. The device was booted using a \gls*{usd} card that loaded the \gls*{pynq} framework. This open-source framework allows developers to program AMD-Xilinx UltraScale+ Zynq \gls*{fpga} devices, such as the one used in this device, using Python. Furthermore, \gls*{pynq} was designed to be used in embedded systems and provides a set of libraries, drivers and Jupyter notebooks to enable easy programmability of \glspl*{fpga} through high-level programming languages like Python. This setup allows for the physical connection and control of the camera feed, which can be streamed in the proposed system.

\subsection{PYNQ Framework}
Booting the device required software to be loaded onto an SD card, in order to leverage the security and parallel hardware execution benefits of the Ultra92-V2 programmable logic, the approach was to use the \gls*{pynq} framework version 2.6 \footnote{Available online, \protect\url{http://www.pynq.io/}, last accessed: 05/05/2022}. This platform features a Linux operating system, along with the Python software package and a Jupyter\footnote{Available online, \protect\url{https://jupyter.org/}, last accessed: 05/05/2022} web server for developing solutions on the board for rapid on-field development and reconfigurability over a network. This image should be flashed onto a \gls*{usd} card with a capacity of at least 16GB and inserted into the board. Once powered on, the board can be accessed by connecting a \gls*{usb} Micro-B cable to a host PC or by setting up a WiFi connection on the local network. The board is configured with the default IP address of 192.168.3.1, which allows access to the locally hosted Jupyter web server.

This revised statement provides more clarity by breaking up the original sentence into several shorter sentences. It also provides more detail on how to access the board, including both USB and WiFi options, and clarifies that the Jupyter web server is hosted locally. Additionally, it uses the passive voice as requested.

\subsection{CUDA}

To develop and execute various components which run on the board, there are prerequisites that should be installed during the setup phase. \gls*{opencv} version 4.5.1\footnote{Available online, \protect\url{https://opencv.org/}, last accessed: 05/05/2022} for Python is used to retrieve frames from an input device, such as a webcam or IP camera. NumPy version 1.16.0\footnote{Available online, \protect\url{https://numpy.org/}, last accessed: 05/05/2022} is also used within the Jupyter notebook to manipulate data structures, such as arrays. This was used to read and write user's credential files. An \gls*{mqtt} broker should also be installed to route the data between publisher and subscribers, so the Mosquitto-\gls*{mqtt} version 1.4.15 \footnote{Available online, \protect\url{https://mosquitto.org/}, last accessed: 05/05/2021} was installed from Ubuntu's open-source universe repository. Finally, to create an \gls*{mqtt} client to publish the stream from the embedded processing system, the Paho-\gls*{mqtt} \footnote{Available online, \protect\url{https://pypi.org/project/paho-mqtt/}, last accessed: 05/05/2022} Python package was installed. At this point, all the prerequisites to develop and execute the system are in place on the \gls*{pynq} Linux environment. A detailed list of the tools and equipment used in the system can be found in Table \ref{tab:equip}.
\begin{longtable}{ |p{3.4cm}|p{2.2cm}|p{4.7cm}| }
\caption{List of equipment and tools}\label{tab:equip}\\
 \hline
 Resource&Type&Description\\
 \hline
 Anvet \gls*{ultra96}&Equipment&System on Chip device for authenticating user key and running the various components of project\\
  \hline
 \gls*{njn}&Equipment&Embedded system capable of achieving higher frame rates and running the various components of project\\
 \hline
 \gls*{rpi4}&Equipment& Low cost embedded system, equipped with 2GB of DDR, ideal for reducing the costs and running the various components of project\\
 \hline
 Full HD 1080p Web Camera&Equipment&Camera used to capture environment for smart camera \\
 \hline
 Vivado \gls*{hls} 2020.1&Development Tool&AMD-Xilinx tool used for creating custom authentication \gls*{ip} core\\
 \hline
 Vivado 2020.1&Development Tool&AMD-Xilinx tool used for creating hardware design and generating the secure bitstream\\
 \hline
 \gls*{pynq}&Framework&Framework for booting the \gls*{ultra96} board with Ubuntu 18.04 and interfacing with the programmable logic\\
 \hline
 OpenCV&Library&Library used to capture input streams, such as camera and online video stream\\
 \hline
 NVIDIA CUDA 11&Development Tool&NVIDIA tool used for accelerating the video streaming using CUDA \gls*{ip} cores\\
 \hline
 NumPy&Library&Various array manipulation functionalities and for saving/reading the user credentials file\\
  \hline
 Paho \gls*{mqtt}&Library&Used to create an \gls*{mqtt} client on the \gls*{ultra96} to publish the camera stream to a topic\\
 \hline
 Mosquitto-\gls*{mqtt}&Software&Used to run a \gls*{mqtt} broker which listens to client publishing to manage and route to the correct subscribers\\
 \hline
 Node-RED&Development Tool&The application layer is created using Node-RED to create a \gls*{iot} flow\\
 \hline
 PuTTY&Terminal Emulator&An emulator which has built in protocols such as SSH to connect to the \gls*{ultra96}\\
 \hline
\end{longtable}

\subsection{Designing the Secure Bitstream of the \gls*{ultra96}}
To secure the system, a bitstream file was created to provide confidentiality through a 256-bit secret key and method of authentication. The key is described in the \gls*{fpga} logic and is embedded within the \gls*{fpga} unit, allowing the system to securely store the private key and prevent it from being exposed in the \gls*{ram}. The system employs a pair of keys, consisting of a public and a private key, to ensure secure message authentication. Nevertheless, the private key is securely stored in the \gls*{fpga} logic, making the proposed method safer than other authentication methods that store private keys externally. The use of a private key stored in the \gls*{fpga} logic ensures that only authorised devices with the corresponding public key are granted access to the system and camera stream. This is achieved through a secure authentication process where the authorised device sends an authentication message encrypted with the public key. The \gls*{fpga} then decrypts the message using the private key stored in its logic to confirm that the device is authorised for providing an additional layer of security to the system, making it more resistant to unauthorised accesses. The high level design flow was used to develop the hardware design at a higher level of abstraction using C/C++ code, which was then converted into optimised \gls*{rtl} code by a compiler. Custom Intellectual Property cores were also included in the design and interfaced with via the \gls*{pynq} framework to run on the \gls*{ultra96} programmable logic. 

\begin{figure}[]

    \centering
    \includegraphics[scale=0.65]{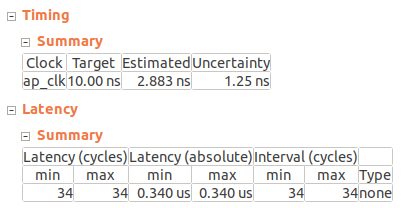}
    \caption{Timing and latency summary. The estimated time is 2.88ns with an uncertainty of 1.25ns, which is below the target clock of 10ns. In terms of latency, both the maximum and minimum values are 34 clock cycles or 0.34us.}
    \label{fig:timing}
\end{figure}

The process of building the authentication \gls*{ip} core begins with using AMD-Xilinx Vivado \gls*{hls} \footnote{Available online, \protect\url{https://www.xilinx.com/products/design-tools/vivado.html}, last accessed: 05/05/2022} software. Using the \gls*{hls} software, the top level function was written in C containing a secret key, authentication method to compare the valid key with the input key, along with the required I/O ports for the \gls*{pl} to interface with the \gls*{ip} block. In this case there were two ports, one of which was the key with a size of 256-bits and the other being the authentication result which was a single bit boolean value to represent whether the input key was valid or not. Due to the size of these ports being relatively small, the AXI4-Lite protocol was utilised for the \gls*{ultra96} processing system to interface with the \gls*{ip} block, as this is generally a suitable design choice for smaller data transfers. To optimise the design, various loop and array optimisation directives were tested in the hope to reduce the estimated clock time and maximum clock cycles. By using the pipeline \textit{Pragma} in the loop to compare the keys, the maximum clock cycles was reduced from 64 to 34. This directive works by reducing the initiation interval for the loop by allowing concurrent execution of the operations. To verify the output of the top level function a test bench was written, this test bench was used by the \gls*{hls} tool during C simulation, synthesis and C/\gls*{rtl} co-simulation to validate that the produced \gls*{rtl} was functionally identical to the C code that was written and therefore confirming that the \gls*{ip} is working as intended to be packed and exported. The timing and latency summary for the \gls*{ip} is presented in Figure \ref{fig:timing}.

\begin{figure}[]
    \centering
    \includegraphics[scale=0.3]{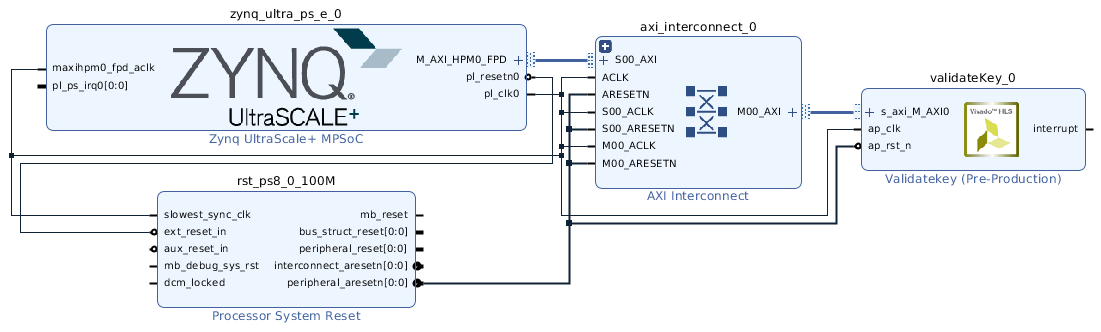}
    \caption{\gls*{ip} Integrator Block Design}
    \label{fig:ip}
\end{figure}

To verify the functionality of the custom \gls*{ip} block, a test bench was written in C. This test bench was used by the \gls*{hls} tool during C simulation, synthesis, and C/\gls*{rtl} co-simulation to validate that the produced \gls*{rtl} was functionally identical to the C code, and therefore confirm that the \gls*{ip} was working as intended. Once the custom \gls*{ip} block was exported, it could be used as part of the wider system by importing it into the AMD-Xilinx Vivado Design Suite. This tool has an \gls*{ip} Integrator, which was used to build the hardware design by integrating the custom \gls*{ip} block with \glspl*{ip} available in the AMD-Xilinx's \gls*{ip} catalogue. A block diagram, shown in Figure \ref{fig:ip}, was generated using the Vivado Tool, containing the Zynq UltraScale+ \gls*{mpsoc} block, which represents the processor of the \gls*{ultra96} and configures clocks, peripherals, and other settings. To transfer the authentication data between the PS and the custom \gls*{ip}, a single memory-mapped \gls*{axi} master and \gls*{axi} slave Interconnect was included. The reset signals were handled by the Processor System Reset \gls*{ip} block.

The custom \gls*{ip} block was integrated into the wider system using the AMD-Xilinx Vivado Design Suite. The \gls*{ip} Integrator tool was used to build the hardware design by combining the created \gls*{ip} block with \glspl*{ip} available in the AMD-Xilinx catalogue. The design was then simulated, synthesised, and implemented to generate a bitstream. The bitstream, .hwh file, and driver file were transferred to the \gls*{ultra96} device to be imported using the \gls*{pynq} Overlay class. This process allows for the custom \gls*{ip} block to be used in the system to provide secure authentication. The \gls*{ip} block is lightweight and only uses a small portion of the programmable logic resources on the device.

\begin{figure}[H]
\hspace*{-1cm} 
    \centering
    \includegraphics[scale=0.5]{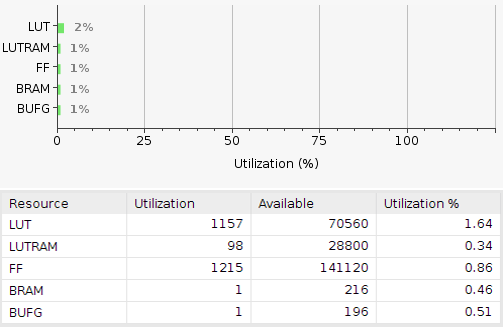}
    \caption{\gls*{ip} Resource utilisation}
    \label{fig:ip_resource}
\end{figure}

The \gls*{ip} block is lightweight and only uses a small portion of the programmable logic resources on the device. Figure \ref{fig:ip_resource} shows the resource utilisation.

\gls*{pynq} is accessed through a local Jupyter web server at 192.168.3.1. It allows the execution of Python packages and libraries on the \gls*{ultra96} board. The \gls*{pynq} Overlay class can be used to view and interface with the PL of the \gls*{ultra96} using the previously created bitstream and default overlay driver to access the IP's ports configured in the drivers file. The authentication result is retrieved using three specific addresses: the start control signal (0x000), the offset of the input key port (0x080), and the offset of the data out port (0x100). The user's symmetric key can be loaded into the input key port a 4-byte integer at a time. The start control signal is set to high to start the \gls*{ip} and the authentication output is read from the data out port. If the key is valid, the camera is initialised and published to the \gls*{mqtt} broker.

\begin{figure}[H]
    \centering
    \includegraphics[scale=0.5]{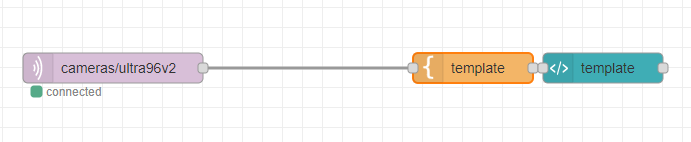}
    \caption{Node-RED Flow. The streaming node that will be enabled only after a successful authentication process.}
    \label{fig:node_red_flow}
\end{figure}

The \gls*{mqtt} broker was configured to automatically start on boot and run on localhost with the port 1883. \gls*{opencv} and Paho-\gls*{mqtt} were also imported for use in capturing and publishing the camera stream. A configuration file was placed on the device to define system parameters such as camera settings, \gls*{mqtt} settings, and the path to the credentials file. This file allows the user to easily update system parameters without requiring knowledge of the system. The data visualisation at the application layer was the final component for the project. This layer is responsible for connecting the clients or subscribers to the \gls*{mqtt} broker and displaying the secured camera feed. Node-RED is a browser-based editor, where flows can be built using a catalogue of nodes to fit custom \gls*{iot} requirements. Additional nodes can also be installed via node package manager. This tool was chosen for the project due to being open source and high productivity, where additional nodes can be quickly inserted into the flow and deployed instantly. The flow that was designed consisted of an \gls*{mqtt} input node, which is configured to the \gls*{mqtt} broker running on the host workstation. This node receives the message payload from the broker as a base64 string and then passes this into an HTML image template, which is finally connected to a dashboard widget template where it is displayed automatically. Once this flow is deployed, the dashboard can be accessed on the local network at the URL 127.0.0.1:1880/ui. The designed flow is shown in Figure \ref{fig:node_red_flow}. 

\section{Results}\label{Ch:results}
The aim of this project was to build a flexible and reconfigurable edge device that could protect the integrity of data within a surveillance and monitoring \gls*{iot} system. To achieve this, a secure authentication mechanism was implemented to guarantee that the edge device could only publish the camera stream when data integrity and authenticity could be assured. This was accomplished by concealing a 256-bit secret key and method of authentication inside a bitstream file, which is the hardware description of an \gls*{fpga} and is difficult to reverse engineer due to being a stream of bits that only describe the hardware logic itself. This provided the necessary confidentiality to protect the key.

The proposed system was tested under the same conditions on the \gls*{ultra96}, \gls*{njn} and \gls*{rpi4} to ensure that each version of the \gls*{iot} device delivered the expected functionality and behaviour. Several tests focused on the integration of the \gls*{ip} within the overall system were carried out to ensure that the system is working as intended. This involved testing the end-to-end process of capturing the camera stream, publishing the data over \gls*{mqtt}, and displaying the stream on the dashboard. These tests were carried out by setting up the \gls*{ultra96} board, \gls*{njn} and \gls*{rpi4} with the boot image and prerequisites, importing the bitstream, and running the python script to capture and publish the camera data. The Node-RED flow was then set up, and the dashboard accessed to verify that the stream was being displayed as expected on all devices. Additionally, the system was tested under various scenarios such as using the correct key, using an incorrect key, and attempting to access the stream without providing a key, to ensure that the system was functioning as intended and that the authorisation process (see Figure~\ref{fig:auth_ip_unit_tests}). was secure (see Figure \ref{fig:auth_ip_unit_tests}). These tests were important to ensure that the \gls*{ultra96}, \gls*{njn} and \gls*{rpi4} provide optimal solution, balanced frame rate and secure key storage. It is clear from Table~\ref{tab:result1} that the hardware implementation (i.e. \gls*{ultra96}) had the same performance has the software implementation (i.e. \gls*{njn} and \gls*{rpi4}) but without exposing the private key into the \gls*{ram}.

\begin{figure}[]
    \centering
    \includegraphics[scale=0.6]{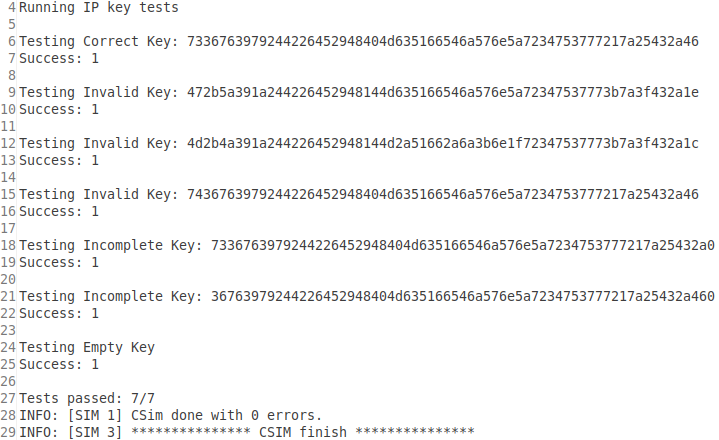}
    \caption{Authentication Unit Tests. The output of the seven unit tests, where the first test involved providing the correct key followed by three tests with incorrect keys, and the last three tests involved two incomplete keys.}
    \label{fig:auth_ip_unit_tests}
\end{figure}

\gls*{ultra96}, the top level function was written in C, containing the secret key and authentication method to compare the valid key with the input key, along with the required I/O ports for the \gls*{ps} to interface with the \gls*{ip} block. The \gls*{axi4}-Lite protocol was used for the \gls*{ultra96} processing system to interface with the \gls*{ip} block, as it is suitable for smaller data transfers. To optimize the design, various loop and array optimisation directives were tested to reduce the estimated clock time and maximum clock cycles. By using the pipeline \textit{Pragma} in the loop to compare the keys, the maximum clock cycles was reduced from 64 to 34.

\begin{table}[htb!]
\caption{Authentication process (number of attempts) per each processing system} \label{tab:result1}
\centering
\begin{tabular}{lcccccc}\hline
Tests          & \multicolumn{2}{l}{\gls*{rpi4}} & \multicolumn{2}{l}{\gls*{njn}} & \multicolumn{2}{l}{\gls*{ultra96}} \\\hline \hline
Correct Key    &        10     &           &    10        &            &       10       &           \\
Invalid Key    &     5        &            &       5     &            &      5        &          \\
Incomplete Key &       7      &            &       7     &            &      7        &             \\
Empty Key      &        4     &            &      4      &            &          4    &            \\ 
Wrong key      &       6      &            &       6     &            &       6       &            \\\hline 
Successful authentications   &      10       &            &     10       &            &      10     &  \\
Unsuccessful authentications   &      22       &            &     22       &            &      22     &  \\\hline           
\end{tabular}
\end{table}

To verify the output of the top level function, in the \gls*{ultra96}, a test bench was written and used by the \gls*{hls} tool during C simulation, synthesis, and C/\gls*{rtl} co-simulation to validate that the produced \gls*{rtl} was functionally identical to the C code and that the \gls*{ip} was working as intended. Once the custom \gls*{ip} block was exported, it could be used as part of the wider system by importing it into the AMD-Xilinx Vivado Design Suite. This tool has an \gls*{ip} Integrator, which was used to build the hardware design by integrating the created \gls*{ip} block with \glspl*{ip} from the AMD-Xilinx's \gls*{ip} catalogue. A single memory-mapped \gls*{axi} master and \gls*{axi} slave Interconnect was included to ensure that the system was able to handle various scenarios that may occur during operation. These tests included different combinations of correct and incorrect keys, as well as edge cases such as missing or incorrect bytes in the key. It was essential that all of these tests passed in order to consider the \gls*{ip} secure and fit for purpose. In addition to these unit tests, the overall system was also tested to ensure that it functioned as intended. This included testing the \gls*{mqtt} communication protocol, the \gls*{pynq} framework, and the Node-RED dashboard. Overall, the testing of the system showed that the objective of building a flexible and reconfigurable edge device was met, as the system was able to securely store and use a secret key and authentication method, and was also easily configurable and adjustable through the use of a configuration file and various software tools. A demo video \cite{murray2023} is available on YouTube\footnote{Available online, \protect\url{https://youtu.be/8AXlf6tRZyo}, last accessed 07/01/2023} demonstrating the system working.
\begin{figure}[H]
    \centering
    \includegraphics[scale=0.6]{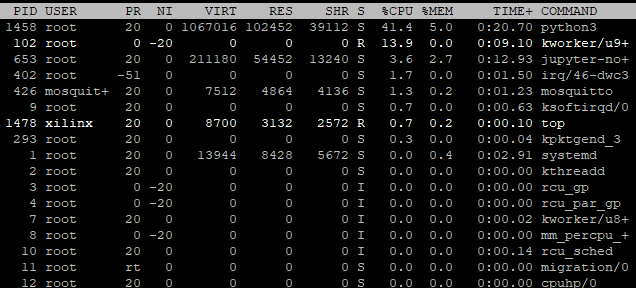}
    \caption{\gls*{ultra96} \protect\gls*{cpu} resource consumption}
    \label{fig:auth_ip_unit_tests}
\end{figure}

The final results obtained for all the devices are listed in Table \ref{tab:results}.
\begin{table}[!hbt]\caption{Results comparison} \label{tab:results}
\resizebox{12cm}{!}{\begin{tabular}{|l|l|c|c|c|l|}
\hline
Device & Security aspects & \multicolumn{1}{l|}{Frame rate no auth} & \multicolumn{1}{l|}{Frame rate with auth} & \multicolumn{1}{l|}{Image size} & Average price \\ \hline
Ultra96 & \begin{tabular}[c]{@{}l@{}}Safest because the private key is \\ never exposed to memory.\end{tabular} & \begin{tabular}[c]{@{}c@{}}14\\ near real-time\end{tabular} & \begin{tabular}[c]{@{}c@{}}14\\ near real-time\end{tabular} & $1920\times1080$                & £300.29 GBP\footnote{Available online, \protect\url{https://uk.farnell.com/avnet/aes-ultra96-v2-g/sbc-arm-cortex-a53-cortex-r5/dp/3050481}, last accessed 15/01/2023}   \\ \hline
\gls*{rpi4} & \begin{tabular}[c]{@{}l@{}}Unsafe because the secure key was\\ exposed to the memory.\end{tabular}    & \begin{tabular}[c]{@{}c@{}}6\\ far real-time\end{tabular}& \begin{tabular}[c]{@{}c@{}}7\\ far real-time\end{tabular}   & $1920\times1080$                & £52.00 GBP\footnote{Available online, \protect\url{https://thepihut.com/products/raspberry-pi-4-model-b?variant=20064052674622}, last accessed 15/01/2023}   \\ \hline
\gls*{njn}  & \begin{tabular}[c]{@{}l@{}}Unsafe because the secure key was\\ exposed to the memory.\end{tabular}    & \begin{tabular}[c]{@{}c@{}}30\\ real-time\end{tabular} & \begin{tabular}[c]{@{}c@{}}30\\ real-time\end{tabular}     & $1920\times1080$                & £61.00 GBP\footnote{Available online, \protect\url{https://thepihut.com/products/nvidia-jetson-nano-2gb-developer-kit}, last accessed 15/01/2023}   \\ \hline
\end{tabular}}
\end{table}

The proposed system was tested under the same conditions on the \gls*{ultra96}, \gls*{njn} and \gls*{rpi4} to evaluate the performance and security of the \gls*{iot} device. The results showed that the \gls*{njn} achieved the highest frame rate of 30 \gls*{fps} (real-time), making it the best in terms of frame rate. The \gls*{rpi4} offered a more cost-effective solution but with a lower frame rate of 6 fps, making it the worst in terms of frame rate. And the \gls*{ultra96} achieved a frame rate of 14 \gls*{fps} and offered a safer solution by securely storing the \gls*{rsa} encryption key in the \gls*{fpga} unit, making it the best in terms of security and performance. 

\section{Discussion and Future Work}\label{Ch:concl_future}
In comparison to other authentication systems that rely solely on the use of a \gls*{cpu}, the proposed \gls*{iot} system utilizing an \gls*{fpga} has several advantages. One main advantage is the improved security provided by storing the secret key and authentication method within the \gls*{fpga} bitstream, as it is not accessible in a readable format outside the device. This is in contrast to a \gls*{cpu}-only system where the secret key and authentication method may be stored in plaintext or encrypted in memory, which could potentially be accessed by an attacker with the appropriate tools and knowledge. On the \gls*{cpu} side, private keys are often stored in the file \$HOME/.ssh/id\_rsa as plain text, which poses a security risk. However, tools like valgrind\footnote{Available online, \protect\url{https://valgrind.org/}, last accessed: 25/03/2023} and Hex editors can be utilised to inspect processes loaded into memory, making it possible to detect any potential breaches. Additionally, one can enhance security by adding authorised public keys to \$HOME/.ssh/authorized\_keys. Nevertheless, in the case of \gls*{fpga} logic, keeping authorised public and private keys described within the logic makes it exceptionally challenging to breach security. This is because \glspl*{fpga} are configured with hardware-level security features that can help to prevent unauthorised access and ensure that the keys remain secure. Although there are risks associated with storing private keys in plain text on the \gls*{cpu} side, there are also measures that can be taken to mitigate these risks, and the use of \gls*{fpga} logic can provide an additional layer of security.

Another advantage of the proposed system is its reconfigurability and flexibility. With the use of a configuration file, the user is able to easily update the location of the bitstream, as well as change the camera input and \gls*{mqtt} settings without requiring in-depth knowledge of the system. This is not necessarily possible with a \gls*{cpu}-only system, as changes to the system may require modifications to the codebase and potentially require the expertise of a software developer. Therefore, the user can decide which programming system to use based on the budget, security and frame rate restrictions. In terms of performance, the \gls*{ultra96} is able to stream the camera feed at a maximum of 14 \gls*{fps}, which is higher than the aim of 6 \gls*{fps} offered by the \gls*{rpi4}. This is due to the efficient resource utilisation of the \gls*{fpga}, as seen in Figure \ref{fig:auth_ip_unit_tests} where the device is only utilizing 63.6\% of its \gls*{cpu} resources. In comparison, a \gls*{cpu}-only system may struggle to handle the processing demands of the camera streaming and \gls*{mqtt} communication simultaneously, potentially resulting in lower frame rates or slower performance.

In summary, the proposed \gls*{iot} system utilising an \gls*{fpga} for authentication offers improved security, reconfigurability, and performance compared to systems that rely solely on a \gls*{cpu}. There are many directions in which future work on this project could go. One possible avenue of research is to improve the security of the system by implementing more advanced forms of authentication. For example, instead of using a simple symmetric key, a more secure method such as a public-private key pair could be used. This would require the use of a cryptographic accelerator or hardware security module to ensure that the key operations can be performed quickly and efficiently on the edge device. Another possibility is to incorporate additional security measures to protect against physical tampering with the device. This could include the use of tamper-evident seals or hardware-based intrusion detection to alert the user if the device has been opened or tampered with.

Another area for improvement could be to optimize the system for better resource utilisation. This could involve using more advanced optimisation techniques during the design phase, or implementing more efficient protocols for communication between the different components of the system. Finally, it would be interesting to explore the possibility of implementing machine learning algorithms on the edge device to enable more advanced forms of data analysis and decision-making. This could involve training a model on the device to identify certain patterns or characteristics in the data, and then using this model to make decisions about how to handle the data. Overall, there are many exciting directions in which this project could be taken, and we believe that it has the potential to make a significant impact in the field of edge computing and \gls*{iot} security.

\section*{Acknowledgements}
The authors would like to express their gratitude to Mr Flemming Christensen and Sundance Multiprocessor Technology for their invaluable support and assistance in the form of AMD-Xilinx training.

\section*{Declarations}
The manuscript has no associated data and the source code can be made available upon request.

\newpage

\bibliography{references}

\end{document}

%% file: glossary.tex
\newacronym{iot}{IoT}{Internet-of-Things}
\newacronym{mpsoc}{MPSoC}{Multi-Processor System-On-Chip}
\newacronym{cpu}{CPU}{Central Processor Unit}
\newacronym{gpu}{GPU}{Graphics Processor Unit}
\newacronym{fpga}{FPGA}{Field-Programmable Gate Array}
\newacronym{pl}{PL}{Programmable Logic}
\newacronym{ram}{RAM}{Random Access Memory}
\newacronym{ic}{IC}{Integrated Circuit}
\newacronym{http}{HTTP}{Hypertext Transfer Protocol}
\newacronym{coap}{CoAP}{Constrained Application Protocol}
\newacronym{mqtt}{MQTT}{Message Queueing Telemetry Transport}
\newacronym{xmpp}{XMPP}{Extensive Messaging and Presence Protocol}
\newacronym{xml}{XML}{Extensible Markup Language}
\newacronym{qos}{QoS}{Quality of Service}
\newacronym{vhdl}{VHDL}{Very High Speed Integrated Circuit Hardware Description Language}
\newacronym{soc}{SoC}{System-on-Chip}
\newacronym{hdl}{HDL}{Hardware Description Languages}
\newacronym{hls}{HLS}{High Level Synthesis}
\newacronym{arm}{ARM}{Advanced Reduced Instruction Set Computing Machine}
\newacronym{rot}{RoT}{Root of Trust}
\newacronym{asic}{ASIC}{Application Specific Integrated Circuit}
\newacronym{rtl}{RTL}{Register Transfer Level}
\newacronym{opencl}{OpenCL}{Open Computer Language}
\newacronym{opencv}{OpenCv}{Open Computer Vision}
\newacronym{usb}{USB}{Universal Serial Bus}
\newacronym{axi4}{AXI4}{ARM eXtensible Interface 4}
\newacronym{axi}{AXI}{ARM eXtensible Interface}
\newacronym{ps}{PS}{Processor System}
\newacronym{ip}{IP}{Intellectual Property}
\newacronym{usd}{uSD}{Micro Secure Digital}
\newacronym{fps}{fps}{frames per Second}
\newacronym{njn}{NJN}{NVIDIA Jetson Nano}
\newacronym{rpi4}{RPI4}{Raspberry Pi 4}
\newacronym{ultra96}{Ultra96}{Ultra96-V2 Development Board}
\newacronym{njpsdk}{NJPSDK}{NVIDIA JetPack SDK}
\newacronym{pynq}{PYNQ}{Python productivity for Zynq}
\newacronym{lora}{LoRa}{Low Range}
\newacronym{lorawan}{LoRaWAN}{Low Range Wide Area Network}
\newacronym{lpwan}{LPWAN}{Low-Power Wide Area Network}
\newacronym{rsa}{RSA}{Rivest–Shamir–Adleman}